\documentstyle [aps,prd,epsf]{revtex}
\input epsf
\begin{document}

\newcommand{\alp}{$\alpha\,\,$}
\newcommand{\xe}{$x_e\;$}
\newcommand{\tdot}{$\dot{\tau}\;\,$}
\newcommand{\be}{\begin{equation}}
\newcommand{\ee}{\end{equation}}
\newcommand{\obh}{$\Omega_B h^2\;$}
\newcommand{\omh}{$\Omega_m h^2\;$}
\newcommand{\och}{$\Omega_c h^2\;$}		
\newcommand{\okh}{$\Omega_K h^2\;$}	
\newcommand{\olh}{$\Omega_\Lambda h^2\;$}
\newcommand{\deln}{$\Delta N_\nu\;$}
	
\title{Recent CMB Observations and the Ionization History of the Universe}
\author{Steen Hannestad$^{1}$ and Robert J. Scherrer$^{2,3}$}
\address{$^1$NORDITA, Blegdamsvej 17, DK-2100 Copenhagen, Denmark}
\address{$^2$Department of Physics, The Ohio State University,
Columbus, OH~~43210}
\address{$^3$Department of Astronomy, The Ohio State University,
Columbus, OH~~43210}
\date{\today}
\maketitle

\begin{abstract}
Interest in non-standard recombination scenarios has been spurred by
recent cosmic microwave background (CMB) 
results from BOOMERANG and MAXIMA, which show an unexpectedly low
second acoustic peak, resulting in a best-fit baryon density
that is 50\% larger than the prediction of big-bang nucleosynthesis (BBN).
This apparent discrepancy can be avoided if the universe has a non-standard
ionization history in which the recombination of hydrogen is 
significantly delayed relative to the standard model. 
While future CMB observations may eliminate this discrepancy, it
is useful to develop a general framework for analyzing non-standard
ionization histories.  We develop such a framework,
examining non-standard models in which
the hydrogen binding energy $E_b$ and the overall expression for
the time rate of change of the ionized fraction of electrons
are multiplied by arbitrary factors. This set of models
includes a number of previously-proposed models as special
cases. We find a wide range of models with delayed recombination that are
able to fit the CMB data with a baryon density in accordance
with BBN, but
there are even allowed
models with {\it earlier} recombination than in the standard
model.
A generic prediction of these models is that the third
acoustic CMB peak should be very low relative to what is
found in the standard model. This is the case even for the models
with earlier recombination than in the standard model, because
here the third peak is lowered by an increased diffusion damping
at recombination relative to the standard model.
Interestingly, the specific
height of the third peak depends sensitively on the model
parameters, so that future CMB measurements will be able
to distinguish between different non-standard recombination
scenarios.
\end{abstract}


\section{INTRODUCTION}

In the past year, observations of the
cosmic microwave background (CMB) fluctuations
by the BOOMERANG \cite{boom} and MAXIMA \cite{max} experiments
have produced data of unprecedented precision on CMB fluctuations 
at small angular scales.  While generally confirming the
adiabatic, flat ($\Omega \approx 1$) model predicted by inflation,
these observations have several puzzling features.  In particular,
the position of the first acoustic peak is at a slightly
larger angular scale than is predicted in the flat model,
and the amplitude of the second peak is unexpectedly low
(see, e.g., Ref. \cite{white}).  If these results are fit
using the standard set of cosmological parameters, the result
is a CMB prediction for the baryon density of
$\Omega_b h^2 \sim 0.03$ \cite{Lange} - \cite{Tegmark}.
In contrast, the prediction for $\Omega_b h^2$ from Big-Bang
nucleosynthesis is $\Omega_b h^2 \sim 0.02$ \cite{OSW,BNT}.

This apparent discrepancy could easily vanish in the light of future
CMB measurements.  For the time being, however, it has led to a great
deal of interest in models with non-standard ionization histories, since
one way to explain the CMB observations and preserve
agreement with the BBN baryon density is to postulate that the
epoch of recombination was delayed to a lower redshift than in
the standard model.  Peebles, Seager, and Hu suggested that
this could occur if there were sources of Ly $\alpha$ photons
present at the epoch of recombination \cite{peebles}.
A more speculative mechanism is a time-variation in the fine-structure
constant, $\alpha$ \cite{hannestad}-\cite{landau}.  The authors of
Ref. \cite{landau} used such a time-variation (along
with changes in the cosmological parameters) as an example of
how one might model non-standard recombination in general.

Because there are a variety of models with non-standard ionization histories,
we feel that is it worthwhile to try to develop a general framework for
analyzing such models.
Ideally, one would like to investigate the consequences for the CMB
of an arbitary $x_e(z)$, the ionization fraction as a function of redshift.
It is obviously impractical to investigate arbitrary functional forms
for $x_e$.  Instead, we have attempted to parametrize deviations
of $x_e(z)$ from the standard ionization history
in terms of a small number of physically-motivated
parameters.  In particular, we multiply the overall
ionization/recombination rates by a free parameter $a$, and the binding
energies by a parameter $b$.
We then calculate the predictions for the CMB fluctuation
spectrum as a function of these parameters and compare with observations.
Our parametrization is discussed in the next section, along with the results
of modifying the ionization history.
Our conclusions are given in Sec. 3.  Although we have compared our
results with the recent BOOMERANG and MAXIMA experiments, our work
provides a general framework for discussing non-standard recombination
scenarios, and it
can be applied to future observations as well.

\section{NON-STANDARD RECOMBINATION}

Given that one cannot examine all possible recombination histories,
what is the best subset to investigate?  We have attempted to model
a subset of such recombination histories which is physically motivated
and, at least to some extent, reduces to the Peebles et al. model
\cite{peebles} and the time-varying $\alpha$
model \cite{hannestad}-\cite{landau} as special cases.  There are two
possible approaches to modifying the ionization history $x_e(z)$.
We can directly modify this ionization history, or we can alter
the evolution equation for $dx_e/dt$.  We use
the latter approach, but we also examine the relationship between
variations in $dx_e/dt$ and the resulting form for $x_e(z)$.

Consider first the equation for $dx_e/dt$:
\cite{peebles2}-
\cite{jones}:
\be
\label{dxdt}
-\frac{dx_e}{dt}={\cal C}\left[{\cal R}n_px_e^2-\beta(1-x_e)
\exp\left(-\frac{B_1-B_2}{kT}\right)\right], \ee
where ${\cal R}$ is the recombination coefficient, $\beta$ is the ionization 
coefficient, $B_n$ is the binding energy of the $n^{th}$ H-atom level and 
$n_p$ is the sum of free protons and H-atoms. The Peebles correction 
factor (${\cal C}$) accounts for the effect of
the presence of non-thermal Lyman-$\alpha$ resonance photons;
it is defined as 
\be {\cal C}=\frac{1+K\Lambda (1-x_e)}{1+K(\Lambda+
\beta)(1-x_e)}.
\label{peebles}
\ee
In the above, \(K=H^{-1}n_pc^3/8\pi\nu_{12}^3\)
(where $\nu_{12}$ is the Lyman-\alp transition frequency), and
$\Lambda$ is the rate of decay 
of the 2s excited state to the ground state via 2 photons~\cite{spitzer}.
The ratio $\beta/{\cal R}$ is fixed by detailed balance.

We modify the evolution history for $x_e$ as follows.  We introduce
two new parameters, $a$, and $b$, into equation (\ref{dxdt}).
The parameter $a$ multiplies the overall rate for $dx_e/dt$, while
$b$ multiplies all of the binding energies $B_n$.  Then
equation (\ref{dxdt}) becomes:
\be
-\frac{dx_e}{dt}= a {\cal C}\left[{\cal R}n_px_e^2-\beta(1-x_e)
\exp\left(-b \frac{B_1-B_2}{kT}\right)\right].\ee
We take $a$ and $b$ to be constants independent of $z$ (a model
with redshift-dependent $a$ and $b$ would simply take us back
to arbitrary functional behavior for $x_e(z)$).
The case $a = b = 1$ corresponds to the standard model.
In order to be completely self-consistent the recombination rate
for helium should also be changed with varying $a$ and $b$. However,
this is a very small effect and in all our calculations the
recombination history of helium is assumed to follow the standard
model. This assumption has no bearing on any of our conclusions.

Our $a$ and $b$ parameters have a simple physical interpretation.
A change in $a$ alone represents a change in the ionization and
recombination rates which preserves detailed balance at a fixed
binding energy, since the ratio of the ionization to recombination
rate is unchanged.  A change in $b$ alone simply shifts the epoch
of recombination up or down in redshift by a fixed amount.

This particular parametrization has several advantages.
The time-varying $\alpha$ model is basically
a special case of this model \cite{hannestad,kaplinghat}.
In the
Peebles et al. model \cite{peebles} delayed 
recombination arises from additional 
Ly-$\alpha$ resonance photons produced by some unknown
source. We have not directly tested this specific model
in ($a$,$b$)-space, but instead we have examined a related model in which the 
rates for the Ly-$\alpha$ and $2s \to 1s$
transitions to the ground state are reduced by a factor 
$\epsilon$ which is assumed to be constant in time.
In this model the Peebles correction factor is changed relative
to Eq.~(\ref{peebles})
\be
\label{newpeebles}
{\cal C}=\frac{1+K\Lambda (1-x_e)}{1+K(\Lambda+
\beta/\epsilon)(1-x_e)}.
\ee

Although this model is very similar in behaviour to that
of the Peebles et al. model \cite{peebles}, they are not completely equivalent.
However, the exact form of the Peebles et al. model is 
quite speculative, and our purpose 
here has just been to test some of the 
possible ways to alter recombination. The Peebles et al.
model could be mimicked almost exactly 
by a time-dependent $\epsilon$,
but this is an unnecessary complication.

In order to examine the viable region in ($a,b$) parameter space
we have performed a $\chi^2$ analysis on the data from the
BOOMERANG \cite{boom}
and MAXIMA \cite{max} experiments.
Our procedure is to maximize
the likelihood for each point in ($a,b$) space for the following free
parameters: the total matter density, $\Omega_m$, the Hubble
parameter, $H_0$, the spectral index of the primordial power
spectrum, $n_s$, and the overall normalization of the spectrum,
$Q$. We have assumed a flat geometry so that 
$\Omega_\Lambda = 1 - \Omega_m$, and also that reionization
is unimportant (assuming that the optical depth to reionization
is small, $\tau \simeq 0$). The vector of free parameters is then
\begin{equation}
\theta = \{\Omega_m,H_0,n_s,Q\},
\end{equation}
and the likelihood function to be optimized
is 
\begin{equation}
{\cal L} \propto A \exp\left(-\frac{(C_l(\theta)
-C_{l,obs})^2}{\sigma^2(C_l)}
\right).
\end{equation}

\begin{figure}
\begin{center}
\epsfysize=13truecm\epsfbox{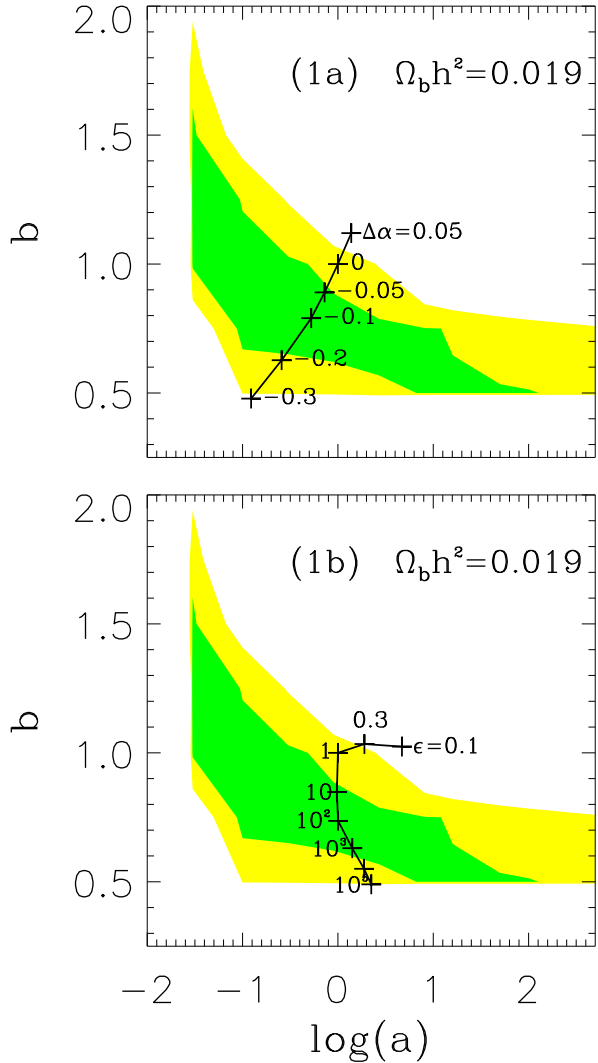}
\end{center}
\vspace*{1.5truecm}
\baselineskip 17pt
\caption{The allowed region in $a,b$ parameter space from the
combined BOOMERANG and MAXIMA data, assuming a baryon density
of $\Omega_b h^2 = 0.019$. The dark shaded (green) region
is the 1$\sigma$ allowed region, and the light shaded (yellow)
is the 2$\sigma$ region. Also shown: 1a) The curve in $a,b$
space which best fits the varying fine-structure
constant ($\alpha$) model for non-standard
recombination; $\Delta \alpha$ is defined as $\Delta \alpha 
\equiv (\alpha-\alpha_0)/\alpha_0$, 1b) The curve which best fits the modified
Peebles et al. model for delayed recombination. The quantity
$\epsilon$ is defined in Eq.~(\protect\ref{newpeebles}).}
\label{fig1a}
\end{figure}
The agreement between our models and the observations, as a function
of $a$ and $b$, is
shown in Figs. 1a and 1b, for $\Omega_b h^2 = 0.019$,
and in Fig. 2 for $\Omega_b h^2 = 0.03$.  In both figures, the area
outside the dark (green) shaded region is excluded at the $1-\sigma$
level, and the area outside the light (yellow) shaded region
is excluded at the $2-\sigma$ level.  The allowed region is a broad
band, in which larger values of $a$ are compensated by smaller
values of $b$.  (Note that the apparent cutoffs at small $a$ and
small $b$ in Fig. 1 are the real boundaries of the confidence region). 
This result hides a great deal of information:
from these graphs alone it is impossible to tell whether the allowed
region corresponds to identical forms for $x_e(z)$ produced
by different values of $a$, $b$, or whether $x_e(z)$ varies
greatly within the allowed region.  The latter is, in fact, the case.
This is shown in Fig. 3, in which we graph the ionization history
for three pairs of $a$, $b$ within the allowed region.  Increasing
$b$ and decreasing $a$ results in a surface of last scattering at
higher redshift which is much broader.

We also display, for comparison, the values of $a$ and $b$
which best fit the models with a change in $\alpha$ (Fig. 1a)
and our version of the Peebles et al. model, given
by equation (\ref{newpeebles}) (Fig. 1b).
The behavior of the time-varying $\alpha$ model is easy
to understand; an increase in $\alpha$ results in an increase
in all of the binding energies (and thus, an increase in $b$)
and it also increases the ionization and recombination rates
(an increase in $a$) \cite{hannestad,kaplinghat}.
Thus, the curve corresponding to time-varying
$\alpha$ runs almost perpendicular to our best-fit contour, and
we find
good agreement for negative values of $\Delta \alpha$,
as in Refs. \cite{battye} - \cite{landau}.  In Fig. 1b,
we see that the correspondence between the modified Peebles, et al.,
model and our $a$, $b$ formalism is more complicated.
The part of the curve corresponding to $\epsilon<1$ corresponds
to faster than normal recombination and is not physically
related to the Peebles et al. model. For $\epsilon>1$ both
$a$ and $b$ are changing with changing $\epsilon$. From Fig.~1
in Ref. \cite{peebles}, it is possible to understand the path taken
by the curve in $a,b$-space. As $\epsilon$ is increased recombination
is pushed to smaller $z$, but at the same time the width of the
recombination surface decreases. Therefore, for increasing
$\epsilon$, the best-fit curve should move down and to the right
in $a,b$ space, exactly what is seen from Fig.~1b. 
This is different from the varying $\alpha$ curve,
where the width of the recombination surface increases as 
$\alpha$ decreases.
This means that the best-fit values in the time-varying $\alpha$
model and the best-fit values in the modified Peebles model lie
in slightly different parts of the $a,b$ plane.
\begin{figure}[t]
\begin{center}
\epsfysize=7truecm\epsfbox{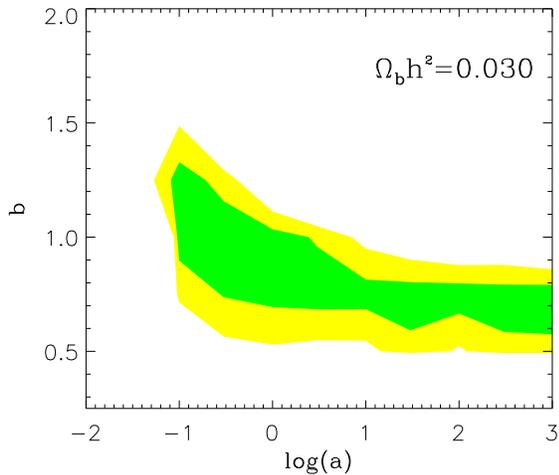}
\vspace*{0.5truecm}
\end{center}
\vspace*{-0.3truecm}
\baselineskip 7pt
\caption{The light (yellow) and dark (green) shaded regions are equivalent to those
in Fig. 1, except that they are calculated assuming a high
baryon density, $\Omega_b h^2 = 0.030$.}
\label{fig1b}
\end{figure}

\begin{figure}[b]
\begin{center}
\epsfysize=7truecm\epsfbox{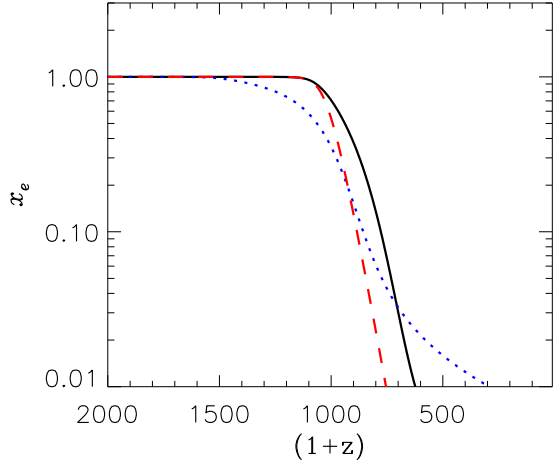}
\vspace*{0.5truecm}
\end{center}
\vspace*{-0.3truecm}
\baselineskip 7pt
\caption{Three different ionization histories, all chosen to lie
within the allowed region. The dotted curve is for 
$a=0.1, b=1.0$, the solid curve for $a=1.0,b=0.75$ and the
dashed for $a=10.0,b=0.65$.}
\label{fig3}
\end{figure}

On first sight the results in Fig. 3
would mean that the large allowed region in
$(a,b)$-space is entirely due to degeneracy between $a$, $b$ and the
other cosmological parameters. In order to investigate this
possibility we have performed a simple Fisher matrix analysis
of how degenerate $a$ and $b$ are with the other cosmological 
parameters. 
The Fisher information matrix is given by \cite{tth}
\be
F_{ij} = \sum_{l=2}^{l_{\rm max}}\frac{1}{\sigma(C_l)^2}
\frac{\partial C_l}{\partial \theta_i}
\frac{\partial C_l}{\partial \theta_j},
\ee
where $i$ and $j$ denote elements in the vector of
cosmological parameters to be determined
and $\sigma(C_l)$ is the uncertainty in the measurement of $C_l$. 
The standard deviation in a measurement of parameter $\theta_i$
is then given by $\sigma(\theta_i)^2 = (F^{-1})_{ii}$. For simplicity
we assume that the measurement error in the $C_l$'s is purely
due to cosmic variance so that
\be
\frac{\sigma(C_l)}{C_l}=\sqrt{\frac{2}{2l+1}}.
\ee
The outcome of this analysis is shown in Figs. 4a and 4b. The precision
with which either $a$ or $b$ can be determined drops considerably if
$a$ and $b$ must be determined simultaneously, 
meaning that $a$ and $b$
are to some extent degenerate. However, the biggest loss of precision
still comes when the other cosmological parameters have to
be determined as well. The partial degeneracy between $a$ and $b$ 
has a simple physical explanation. If the redshift of recombination
is lowered the radiation content at recombination is lower, leading
to a smaller early ISW effect. On the other hand, if the width
of the recombination surface is narrowed, the diffusion damping
of fluctuations is smaller. These two effects can to some extent
compensate each other, and lead to a partial degeneracy between
$a$ and $b$.  However, the degeneracy is not exact, because the
early ISW effect dominates near the first peak in the spectrum,
while the diffusion damping dominates at larger $l$.

\begin{figure}
\begin{center}
\epsfysize=9truecm\epsfbox{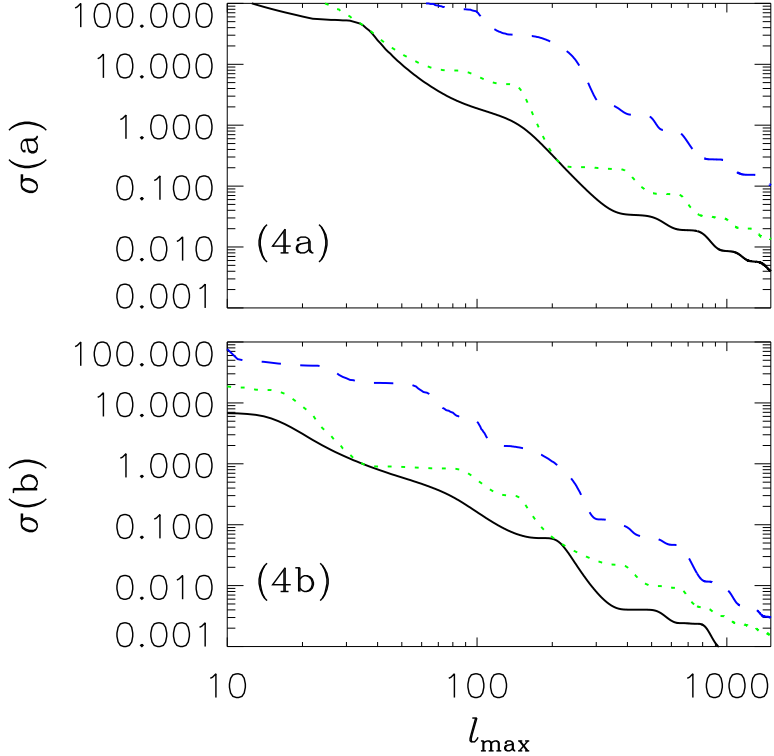}
\end{center}
\vspace*{1.5truecm}
\baselineskip 22pt
\caption{The expected accuracy with which $a$ can be measured, as
a function of the maximum $l$ that the power spectrum can be 
measured to. We have assumed that the power spectrum can be measured
at all $l$, up to $l_{\rm max}$, to within cosmic variance
$\sigma(C_l)/C_l=[2/(2l+1)]^ {1/2}$. Fig. 4a) The solid line is the expected
precision if $a$ alone needs to be determined from the CMB, the dotted
is the case where $a$ and $b$ must be simultaneously determined,
and the dashed the case where all parameters $(\Omega_m,\Omega_b,
H_0,n_s,Q,a,b)$ must be determined.
Fig. 4b) The same as for Fig. 4a, except that $a$ and $b$ are
interchanged.  The fiducial model is $a=b=1$.}
\label{fig4a}
\end{figure}

A second possible approach to the problem of the CMB as a function
of a generic recombination history would be to
modify $x_e(z)$ directly by hand.  This is a somewhat unphysical
approach, so we have instead examined the general behavior of
$x_e(z)$ as a function of $a$ and $b$.
The two key properties of the ionization history which affect
the CMB are the redshift of last-scattering, and the width
of the last-scattering surface.  We parametrize the former
in terms of $z_{1/2}$, the redshift at which $x_e = 0.5$.
To estimate the width of the last scattering surface, we
define the parameter $\Delta z$ to be $z(x_e = 0.9) - z(x_e = 0.1)$.
While both of these definitions are somewhat arbitrary, they
serve the desired purpose of indicating the redshift and width
of the last scattering surface.  The behavior of these
quantities as a function of $a$ and $b$ is illustrated in Figs. 5
and 6, and in Fig. 7 we show $\Delta z/z_{1/2}$ as a function of $a$
and $b$.
\begin{figure}
\begin{center}
\epsfysize=7truecm\epsfbox{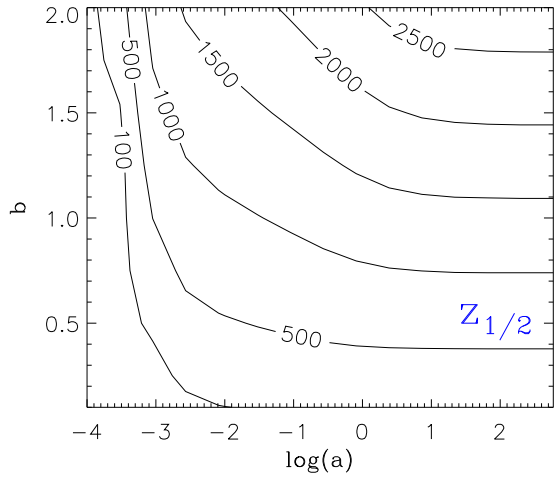}
\vspace*{0.5truecm}
\end{center}
\vspace*{-0.3truecm}
\baselineskip 7pt
\caption{The redshift of last scattering $(z_{1/2})$ as a function
of $a$ and $b$.}
\label{fig5}
\end{figure}
\begin{figure}
\begin{center}
\epsfysize=7truecm\epsfbox{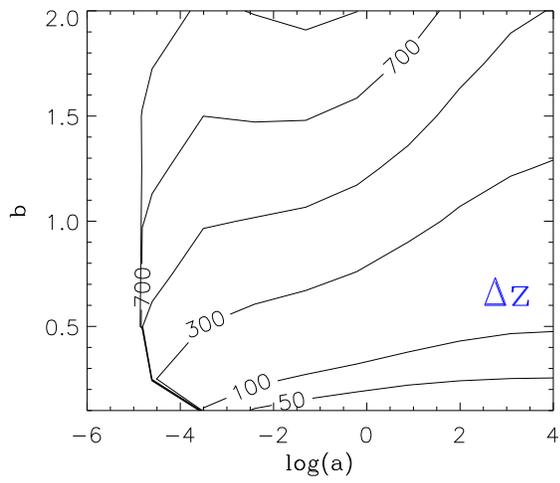}
\vspace*{0.5truecm}
\end{center}
\vspace*{-0.3truecm}
\baselineskip 7pt
\caption{The width of the last scattering surface $(\Delta z)$ as a function
of $a$ and $b$.}
\label{fig6}
\end{figure}
\begin{figure}
\begin{center}
\epsfysize=8truecm\epsfbox{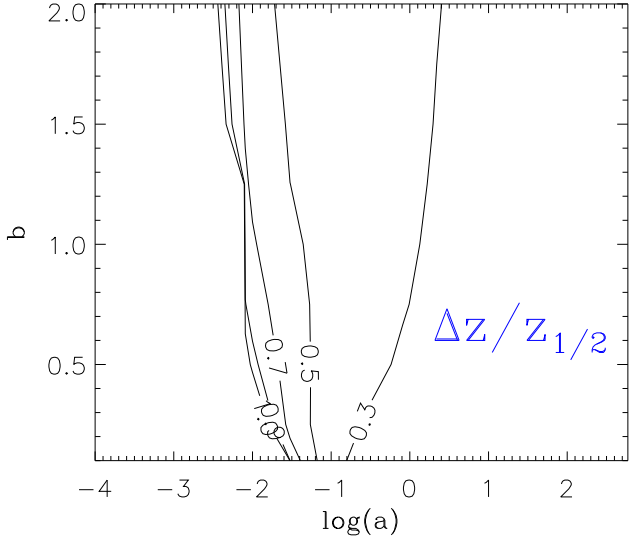}
\vspace*{0.5truecm}
\end{center}
\baselineskip 17pt
\caption{The width of the last scattering surface
relative to the redshift of last scattering, $\Delta z/z_{1/2}$, as a function
of $a$ and $b$.}
\label{fig7}
\end{figure}

Several features are obvious from these plots.  In Fig. 5, we see
that $z_{1/2}$ is a function only of $b$ (and is independent of $a$)
in the limit of large $a$.  We see the same effect in Fig. 6, in which
$\Delta z$ becomes a function only of $b$, but only for the case of small
$b$.
These results follow from the fact
that for large $a$, the ionization fraction tracks its equilibrium value
nearly exactly, and the equilibrium value of $x_e$ at fixed $z$
depends only on the binding energies (and hence, on $b$).  This
argument breaks down for small $a$, since in this case the ionization
fraction no longer tracks the equilibrium abundance.  The argument
breaks down for large $b$ for a more complicated reason.
Roughly speaking, $x_e$ tracks its equilibrium abundance
as long as $(dx_e/dt)/H >1$, where $H$ is the expansion rate.
At large $b$, recombination occurs earlier, when $H$ is
larger (smaller cosmic time), requiring a larger value of $a$
to maintain equilibrium.  This effect is more apparent in Fig. 6
than in Fig. 5 because of the way that we have defined
$z_{1/2}$ and $\Delta z$.  The value of $z_{1/2}$ is
the redshift at which $x_e = 0.5$, while $\Delta z$ depends
on the much later redshift at which $x_e = 0.1$.  Hence,
$z_{1/2}$ will be independent of $a$ as long as equilibrium
is maintained down to the redshift at which $x_e = 0.5$,
while in order for $\Delta z$ to be independent of $a$,
equilibrium must be maintained down to the much lower redshift
at which $x_e = 0.1$.

In the limit of
extremely small $a$, recombination is continuing at the present, and
the lower limit $x_e = 0.1$ we use to calculate $\Delta z$ is never
reached.  Hence, there is a
limiting value for $a$ below which $\Delta z$ is undefined;
this is reflected in the degeneracy of our contours for small
$a$ in Fig. 6.

Finally, if we define the width of our last scattering
surface relative to the redshift of last scattering (as in Fig. 7),
then we see that $\Delta z/z_{1/2}$ is essentially a function of $a$,
and is nearly independent of $b$.  At large
$a$, the ionization fraction tracks its equilibrium value nearly
exactly (and so is independent of $a$).
Furthermore, when $x_e(z)$ is given by the Saha equation,
$\Delta z/z_{1/2}$ is independent of the
binding energy.  Thus $\Delta z/z_{1/2}$ becomes independent of
both $a$ and $b$ for large values of $a$, as seen in Fig. 7.

Figs. 5 and 6 indicate that scanning over all possible values of $a$
and $b$ is nearly equivalent to scanning over all possible
values of the redshift and width of the last scattering surface.
Hence, our method of varying $a$ and $b$ provides a quite
general study of arbitrary ionization histories.  The one
exception is the case of large $z_{1/2}$ and small $\Delta z$,
which does not correspond to any values of $a$ and $b$.  However,
this case corresponds to a decrease in
$x_e$
which occurs faster than for the equilibrium case.  It is difficult
to imagine a mechanism for achieving this.
Furthermore, we expect the $C_l$ spectrum to become independent
of $\Delta z$ for sufficiently small $\Delta z$, since the
thickness of the last-scattering surface will become irrelevant
when it becomes so narrow that diffusion damping can be neglected.
For completeness we have also calculated $\chi^2$ for the case
of $\Delta z = 0$, for which we take $x_e$ to be a step function.
These are shown in Fig. 8. We see that no good fit is obtained
for any value of $z_{1/2}$ ($\chi^2_{\rm min}/{\rm d.o.f} \simeq 3.8$).

If we consider the allowed region in Fig. 1 as a function of
$z_{1/2}$ and $\Delta z$, rather than $a$, and $b$, we find that
both $\Delta z$ and $z_{1/2}$ are restricted by the current
CMB observations.  At the 95\% confidence level, we have
$300 < \Delta z < 900$ and $500 < z_{1/2} < 2000$.  From Fig. 7, we see
that our limits translate into an upper bound on $\Delta z/z_{1/2}$ of
$\Delta z/z_{1/2} < 0.5$.
\begin{figure}
\begin{center}
\epsfysize=8truecm\epsfbox{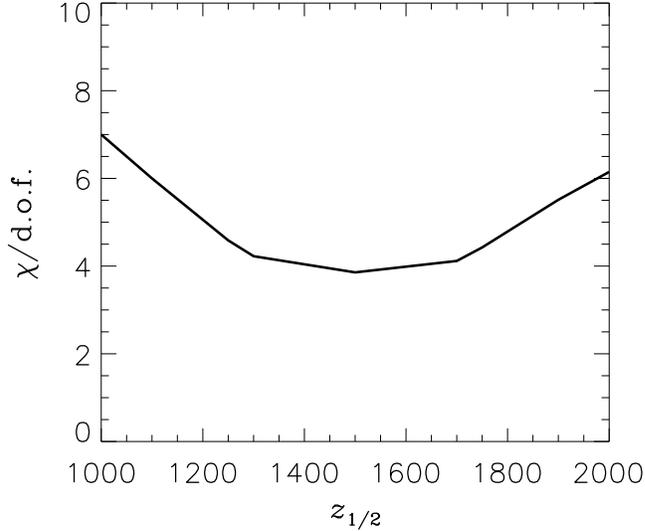}
\end{center}
\vspace*{-0.3truecm}
\baselineskip 7pt
\caption{The $\chi^2$ for the case of $\Delta z = 0$ as a function
of $z_{1/2}$.}
\label{fig8}
\end{figure}
\begin{figure}
\begin{center}
\epsfysize=8truecm\epsfbox{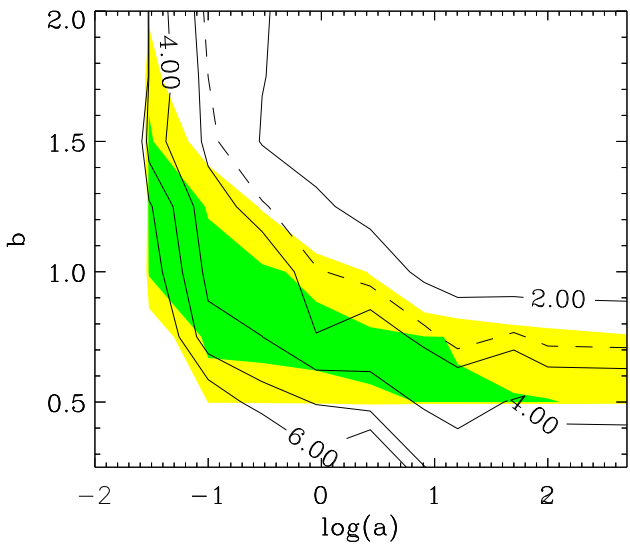}
\end{center}
\vspace*{-0.3truecm}
\baselineskip 7pt
\caption{The height of the first peak relative
to the height of the third peak, $X \equiv C_l({\rm 1st \,\, peak})/
C_l({\rm 3rd \,\, peak})$, as a function of $a$ and $b$.  The dashed
curve gives the standard model ($a=b=1$) value of $X = 2.65$.}
\label{fig9}
\end{figure}
\section{CONCLUSIONS}

Our investigation of the effect of a non-standard ionization history
on the CMB using a change in the overall ionization/recombination
rates (through the parameter
$a$) and in the binding energies (through the parameter $b$) appears
to provide a very general framework for studying such variations
in the standard model.  In particular, it seems possible to model
arbitrary changes in both the epoch of last scattering and in
the width of the last-scattering surface through such variations,
with the exception of ionization histories having a narrow, high-redshift
surface of last scattering, for which the ionization fraction
decreases faster
than in the equilibrium case.

Our results indicate that the BOOMERANG and MAXIMA results with
$\Omega_b h^2 = 0.019$ (from BBN) can be well-fit over a broad
range of choices for $a$ and $b$.
However,
the lower bounds on $a$ and $b$ are quite robust
(i.e., nearly
independent of variation in the other parameter).
We find that  $b >0.5$ and $\log(a) > -1.6$ at the 95\% confidence level.
If we restrict $a$ to
its standard model value, changing only the binding energies,
the allowed (95\%) region is given by $0.5 < b < 1.1$.
Similarly, if we restrict the binding energies to be unchanged
($b = 1$) and change only the overall rates, then $a$ is constrained
to lie in the range $-1.6 < \log(a) < 0.4$.  Our best-fit model for
$\Omega_b h^2 = 0.019$ has a $\chi^2$ per degree of freedom of
0.50.  In comparison, the best fit for standard reionization
($a=b=1$) with $\Omega_b h^2 = 0.030$ gives $\chi^2$ per degree
of freedom of 0.82, so both models are good fits.  (These numbers
include a 10\% calibration error for BOOMERANG and 4\% for
MAXIMA).

One robust prediction of all of these models is
a decrease in
the amplitude of the third peak, relative to its height in the standard
model.  In Fig. 9, we show the ratio $X$ of the height of the first peak
to the height of the third peak as a function of $a$ and $b$.
The standard model ($a=b=1$, $\Omega_b h^2 =0.019$)
value, $X = 2.65$, is shown as a dashed curve.
The entire $1\sigma$ region gives values for $X$ larger than
in the standard model.  In comparison, the best fit model with high baryon
density ($\Omega_b h^2 = 0.030$)
and a standard recombination history produces a much higher third peak
($X = 1.82$) than in the standard model.

Our results are more general than previous
comments that models with late
recombination can
be distinguished from high $\Omega_b h^2$ models
by the amplitude of the third peak.  Our allowed region
includes models with recombination at slightly {\it higher} redshift
than in the standard model (these models lie on the extreme left-hand
side of Fig. 9).  For these
models, the third peak is still reduced in amplitude, but
this reduction is
due to diffusion
damping from an increase in the width of the surface of last scattering,
rather than from a decrease in the redshift of last scattering
as in Refs. \cite{peebles} - \cite{landau}.  Our results
seem to imply that any modification in the recombination history
which fits the BOOMERANG
and MAXIMA observations will result generically in a decrease
in the height of the third peak.

However, we also find that the actual height of the
third peak is a function
of $a$ and $b$.  From Fig. 9
it can be seen that
this ratio increases
as we move through the allowed region from small $a$, large $b$
to large $a$, small $b$.  So if the universe did have a non-standard
ionization history,
the amplitude of the third peak as shown in
Fig. 9 should allow us to determine precisely
the nature of the deviation of $x_e(z)$ from its standard evolution.

\vskip 0.2in
\noindent
We thank U. Seljak and M. Zaldariagga for the use of CMBFAST \cite{CMBFAST}.
R.J.S. was supported in part by the DOE (DE-FG02-91ER40690).

\end{document}